\documentclass[runningheads,a4paper]{llncs}

\usepackage[T1]{fontenc}
\usepackage{graphicx}
\usepackage{algorithm}
\usepackage{algpseudocode}
\usepackage{amsmath}
\usepackage{amssymb}
\usepackage{float}
\usepackage{multirow}
\usepackage{placeins}

\newcommand{\runinhead}[1]{\par\noindent\textbf{#1.}\enspace\ignorespaces}

\begin{document}

\title{Latency-Optimal Adaptive Split Inference for Privacy-Preserving Cloud-Edge-End Collaboration}

\titlerunning{Latency-Optimal Split Inference}

\author{Yi Li\inst{1} \and Peng Zhang\inst{1}\thanks{Corresponding author.} \and Man Ho Au\inst{2}}

\authorrunning{Y. Li et al.}

\institute{Guangdong Provincial Key Laboratory of Intelligent Information Processing,
College of Electronics and Information Engineering, Shenzhen University,
Shenzhen 518060, Guangdong, China\\
\email{stabilelee@foxmail.com, zhangp@szu.edu.cn}
\and Department of Computing, The Hong Kong Polytechnic University, Hong Kong, China\\
\email{man-ho-allen.au@polyu.edu.hk}}

\maketitle

\begin{abstract}
Internet of Things (IoT) end devices are increasingly expected to support privacy-sensitive batch inference, yet their limited computational resources often make full local execution of convolutional neural networks impractical.
This paper presents a latency-optimal adaptive split inference framework for privacy-preserving cloud-edge-end collaboration. 
The end device acts as the trust anchor, executes the plaintext model prefix, encrypts the split activation using fully homomorphic encryption (FHE), and
keeps the secret key locally, while the edge and cloud execute assigned model segments only on FHE ciphertexts. 
We formulate collaborative encrypted inference as a split-pair selection problem over
an end-side split point and an edge-side termination point. 
The proposed planner jointly models plaintext prefix execution, encryption, communication, edge-side FHE execution, and cloud-side FHE completion, and supports both convolution-level and block-level split granularities.
Experiments on CIFAR-10 and PathMNIST show that the proposed
convolution-level collaborative scheme achieves amortized end-to-end speedups of
approximately $12.9{\times}$ over full-cloud FHE and
$3.9{\times}$ over the block-level alternative, while preserving
the corresponding plaintext-model accuracy. Including modeled communication,
the amortized latencies are 1033.279~s/sample on CIFAR-10 and
1023.429~s/sample on PathMNIST.
\end{abstract}

\section{INTRODUCTION}
Internet of Things (IoT) applications increasingly rely on end
devices to acquire and preprocess privacy-sensitive data in settings
such as pathology slide digitization and archival medical image
analysis.
Convolutional neural networks (CNNs) are widely used to provide
visual recognition and decision support for these workloads.
However, IoT end devices are typically constrained by limited computing
capacity, memory, battery supply, and thermal constraints. 
Running the full model locally may therefore lead to long
processing time and high energy consumption.
Directly offloading inference to edge or cloud servers can reduce the
device-side burden, but it exposes raw inputs or semantically rich
intermediate features to external infrastructure.

Split inference distributes CNN execution across multiple tiers by executing an initial prefix on the end device and delegating the remaining layers to edge or cloud servers. Recent studies have shown that this execution model is practical under heterogeneous resource constraints. 
The deep neural network (DNN) partitioning framework DNN
Surgery accelerates edge inference by selecting layer partition points
between the device and edge~\cite{dnnsurgery}, while an adaptive
end-edge-cloud framework further predicts layer-wise latency and divides
DNN computation across all three tiers~\cite{adaptiveeec}.
Latency- and privacy-aware CNN distributed inference further demonstrates that split selection should consider both response time and feature-level privacy in reliable artificial intelligence (AI) systems~\cite{latencyprivacycnn}.
However, conventional split inference falls short for privacy-sensitive IoT applications when intermediate representations are transmitted in plaintext, because these representations may still reveal information about the original input,
user attributes, or the final prediction.

Fully homomorphic encryption (FHE) is a natural cryptographic primitive for this setting because it allows external servers to compute directly on ciphertexts without
observing plaintext inputs or intermediate features. 
Prior encrypted inference systems have demonstrated the feasibility of FHE-compatible
neural network evaluation~\cite{cryptonets,hemet,autofhe}, and homomorphic
encryption has also been adopted in cloud-edge-end privacy-preserving machine learning~\cite{yangceeiot}. Nevertheless, full-cloud FHE deployment remains costly due to ciphertext expansion, encrypted communication, and homomorphic arithmetic. 
Moving the encrypted workload entirely to the edge is also insufficient, since edge servers are closer to the user but have lower computing capacity than the cloud. The central challenge is therefore to determine where to split a CNN and how to divide its encrypted suffix between the edge and cloud tiers so that privacy is preserved while end-to-end latency is minimized.

This paper proposes a privacy-preserving split inference framework for cloud-edge-end collaboration. 
The end device serves as the trust anchor, while the edge and
cloud execute the encrypted suffix according to the latency-minimizing
split pair selected by the planner.
The main contributions of this work are summarized as follows.
\begin{itemize}
\item
We design and integrate a privacy-preserving cloud-edge-end split inference framework for CNNs, where the end device remains the trust anchor by keeping the secret key and all plaintext data locally, and all external computation at the edge and cloud is performed strictly on FHE ciphertexts without ever decrypting intermediate results.

\item We formulate collaborative encrypted inference as a latency minimization problem over split pairs $(q,e)$ and develop a split-granularity-aware planner. The proposed model jointly captures local plaintext execution, encryption, communication, edge-side FHE execution, and cloud-side FHE completion, while supporting both convolution-level and block-level partitioning to avoid bottlenecks in edge-side encrypted computation.

\item
We implement a CKKS/OpenFHE prototype for CNN models designed
for FHE. Experiments on CIFAR-10 and PathMNIST show that convolution-level
collaboration substantially reduces latency relative to both baselines
while matching the corresponding plaintext model's accuracy.
\end{itemize}

\section{RELATED WORK}

\runinhead{Privacy-Preserving Neural Network Inference}
Fully homomorphic encryption has been widely studied as a cryptographic
basis for privacy-preserving neural network inference. CryptoNets
demonstrated that neural networks can be evaluated over encrypted inputs
and returned as encrypted predictions, establishing an early full-cloud
FHE inference paradigm~\cite{cryptonets}. Since FHE schemes support
additions and multiplications but not comparison-based nonlinearities,
subsequent works have redesigned neural networks into FHE-compatible
forms by replacing ReLU and other nonlinear activations with square or
low-degree polynomial approximations~\cite{hemet,autofhe}. Other studies
focus on reducing the large overhead of encrypted inference through
packing, layout optimization, and automated compilation. HyPHEN improves
encrypted inference throughput through hybrid ciphertext packing and
layout optimization~\cite{hyphen}, while Orion provides a recent
framework for deep FHE inference with automated optimization across
encrypted operators~\cite{orion}. Recent work has also pushed practical
FHE-based CNN inference further by combining packing, polynomial
evaluation, and hardware-aware optimization~\cite{practicalfhecnn}.
FHEON provides configurable CKKS/OpenFHE building blocks for
encrypted neural networks and evaluates several CNN families, including
a VGG-16 tailored to CIFAR-10, on a consumer-grade CPU~\cite{fheon}.
Bi-CryptoNets explores a different acceleration
direction by decomposing the input into sensitive and insensitive
segments, processing them through ciphertext and plaintext branches,
respectively~\cite{bicryptonets}. These works show that FHE-compatible
neural inference is feasible and can be accelerated through model,
packing, compiler, or mixed plaintext-ciphertext designs. However, they
primarily focus on encrypted model execution itself, and generally do not
address how an FHE-compatible CNN should be partitioned across end,
edge, and cloud tiers under a unified latency objective.
\runinhead{Split Inference and End-Edge-Cloud Collaboration}
Split inference and collaborative intelligence have been extensively
studied to reduce the latency and resource burden of deep neural network
inference on constrained devices. Neurosurgeon profiles layer-wise
computation and communication costs to decide whether DNN layers should
run on a mobile device or in the cloud~\cite{neurosurgeon}. More recent
surveys and systems studies extend this line toward cloud-edge-end
hierarchies, adaptive partitioning, and privacy-aware split execution~\cite{dnnpartsurvey,saltedinference}. DNN Surgery further studies layer
partitioning under dynamic edge conditions and searches for efficient
partition plans~\cite{dnnsurgery}. More recently, latency- and
privacy-aware CNN distributed inference has considered both service
latency and privacy risk in reliable AI systems~\cite{latencyprivacycnn}.
These approaches establish the importance of model partitioning,
collaborative execution, and split-point search. Nevertheless, most of
them assume plaintext intermediate features or lightweight privacy
metrics rather than FHE-protected remote computation. As a result, their
cost models do not capture encryption latency, ciphertext expansion,
FHE-specific kernel overhead, or the different encrypted execution costs
of edge and cloud tiers. In contrast, this work adopts the
established CKKS scheme without modifying its cryptographic construction
and studies the placement of encrypted CNN computation across end, edge,
and cloud tiers using a unified latency model for the plaintext prefix,
encryption, ciphertext communication, edge-side FHE execution, and
cloud-side FHE completion.

\section{PRELIMINARIES}
This section reviews the cryptographic background of the fully
homomorphic encryption scheme CKKS and the stage-level view of convolutional
neural networks on which the proposed split inference framework is built.

\subsection{CKKS Scheme}\label{sec:prelim_ckks}

\runinhead{Scheme Overview} The Cheon--Kim--Kim--Song (CKKS) scheme~\cite{ckks} supports approximate
arithmetic over encrypted vectors of real numbers and is therefore well
suited to neural network inference. Its plaintext and ciphertext spaces
are defined over the polynomial ring $R_{Q}=\mathbb{Z}_{Q}[X]/(X^{N}+1)$,
where the ring dimension $N$ is a power of two and $Q$ is the ciphertext
modulus. A real-valued message is mapped to a plaintext polynomial after
multiplication by a scaling factor $\Delta$ that controls numerical
precision, so all homomorphic results are recovered up to a bounded
approximation error. Encryption and decryption are performed with a
public/secret key pair, and the secret key is held exclusively by the
trusted end device, so any party that processes ciphertexts without it
does not obtain plaintext information under the CKKS security
assumptions.

\runinhead{Homomorphic Operations} CKKS natively supports homomorphic
addition and multiplication over ciphertexts. Operators outside addition
and multiplication, such as comparison-based nonlinearities, are not
directly available and require special handling, which is addressed in
Section~\ref{sec:split_inference}.

\runinhead{SIMD Packing} CKKS encodes a vector of values into the slots
of a single ciphertext, enabling single-instruction multiple-data
evaluation in which one homomorphic operation is applied to all slots
simultaneously. Slot-wise reduction, as required by fully connected
projections, is realized through ciphertext rotations and the associated
summation routine.

\subsection{Stage-Based Split Inference}\label{sec:prelim_split}

A convolutional neural network can be represented as an ordered sequence
$\mathcal{L}=(\ell_{1},\ell_{2},\ldots,\ell_{M})$, where each stage is a
computation unit such as a convolutional layer, a block boundary, a
pooling transition, or a fully connected layer. The end-to-end inference mapping is
\begin{equation}
\label{eq:stage-composition}
f(\mathbf{x};\Theta)=f_{M}\circ f_{M-1}\circ\cdots\circ f_{1}(\mathbf{x}),
\end{equation}
where $\mathbf{x}$ denotes the input sample, $\Theta$ represents the
network parameters, and $f_m$ is the mapping of stage $\ell_m$. Because
the output feature tensor of one stage is the complete input required by
the next stage, a CNN can be partitioned at stage boundaries without
changing the inference function, provided that the original stage order
and model parameters are preserved. In plaintext distributed inference,
the prefix of the network is executed on one device, the intermediate
feature tensor is transmitted to another device, and the remaining stages
continue from that tensor. This stage-based execution view provides the
basic feasibility of model partitioning and serves as the notation basis
for the cloud-edge-end collaborative framework formalized next.

\section{SYSTEM MODEL AND PROBLEM FORMULATION}\label{sec:system_model}
This section introduces the privacy-preserving cloud-edge-end split inference architecture, states the threat assumptions, and formalizes the latency objective.

\subsection{Inference Architecture}
This work considers a three-tier collaborative inference architecture
comprising end devices, edge servers, and cloud servers. 
Designed to reduce the processing cost of privacy-sensitive encrypted inference, the architecture jointly leverages data locality, resource asymmetry, and remote computing capacity.
Its core principle is that raw user data never leave the end device, while only protected intermediate representations are transmitted for collaborative processing.

\begin{figure}[H]
\centering
\includegraphics[width=0.8\textwidth]{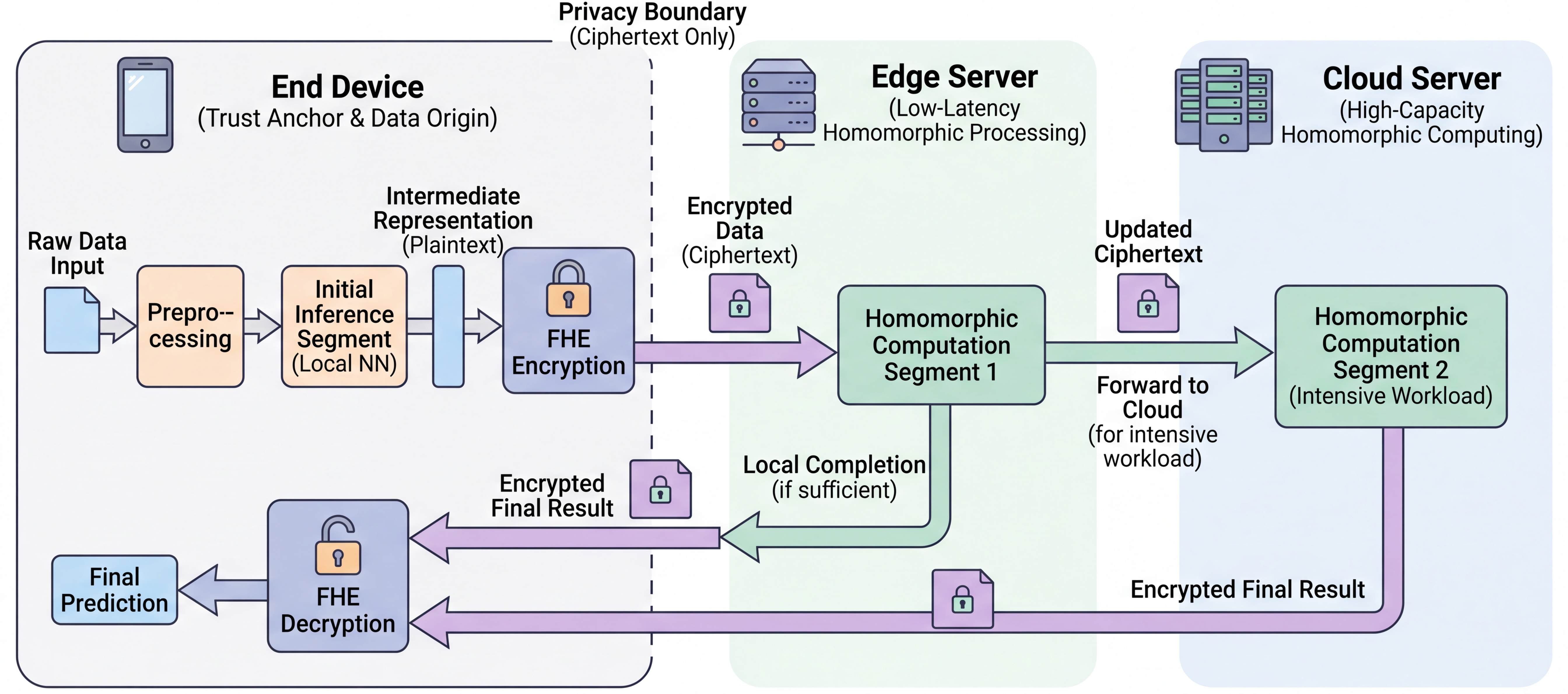}
\caption{System architecture of the privacy-preserving cloud-edge-end split inference.}
\label{fig:ceec_system_architecture}
\end{figure}

\runinhead{End Device} 
As the data origin and primary trust anchor, the end device acquires and preprocesses input and executes the initial segment of the neural inference task locally. 
Before any outward transmission, it encrypts the intermediate representation using an FHE scheme. After remote homomorphic computation is completed, it receives the encrypted result, decrypts it locally, and outputs the final prediction. 

\runinhead{Edge Server}
Deployed close to the end device in the conceptual
architecture, the edge server acts as the first external processing tier
and evaluates its assigned model segment directly on FHE ciphertexts.
Upon receiving the encrypted data, the edge performs a continuous
segment of homomorphic computation and then either forwards the updated
ciphertext to the cloud or returns the final encrypted output to the end
device when the selected plan terminates at the edge. 

\runinhead{Cloud Server} 
The cloud server provides the high-capacity computing tier. It completes the remaining inference workload that is more efficiently assigned to the cloud due to computational intensity, service-scale demand, or end-to-end latency considerations.
Critically, the cloud participates only in FHE-domain processing and requires no access to raw inputs, plaintext intermediate features, or the private decryption key.

\subsection{Threat Model}
In the threat model considered in this work, end devices are assumed to
operate within a trusted security domain, whereas edge and cloud
servers are located outside this trusted boundary. All participating
entities are assumed to be authenticated and to execute the
cloud-edge-end collaborative inference protocol faithfully. However, the
edge and cloud are regarded as honest-but-curious adversaries: although
they do not arbitrarily deviate from the prescribed protocol, they may
attempt to infer users' private information from the data available to
them during distributed inference. In particular, a semi-honest edge or
cloud server may analyze transmitted protected states or encrypted
outputs in an effort to reconstruct the original input, infer sensitive
feature information, or deduce the final prediction outcome. The privacy
boundary does not rely on any non-collusion assumption: even if the edge
and cloud combine their ciphertext observations, no external tier
receives the secret key. Under
these assumptions, the principal security objective of the proposed
framework is to prevent the leakage of raw data, plaintext intermediate
representations, and final plaintext predictions before authorized local
decryption.

\subsection{Problem Formulation}
We formulate the privacy-preserving cloud-edge-end split inference as a constrained decision problem over a three-tier execution path. 
Let the CNN be represented as an ordered stage sequence $\mathcal{L} = (\ell_{1},\ell_{2},\ldots,\ell_{M})$.
The system must determine an end-side split point $q$ and an edge-side
termination point $e$. 
Under a feasible decision tuple $(q,e)$, stages $\ell_{1}$ to $\ell_{q}$
are executed locally in the plaintext domain, stages $\ell_{q+1}$ to $\ell_{e}$ are processed at the selected edge server over protected intermediate data, and stages
$\ell_{e+1}$ to $\ell_{M}$ are completed by the cloud when $e<M$.

The end-to-end service latency of a decision tuple is jointly determined by local prefix inference, transformation of the split activation into a protected representation, inter-tier communication, protected execution at the edge, and cloud-side completion of the residual suffix when needed. Let $T_{\mathrm{tot}}(q,e)$ denote the total latency, which can be expressed as:
\begin{equation}
\label{eq:total-latency}
T_{\mathrm{tot}}(q,e)=T_{\mathrm{end}}(q)+T_{\mathrm{enc}}(q)+T_{\mathrm{comm}}(q,e)+T_{\mathrm{edge}}(q,e)+T_{\mathrm{cloud}}(e),
\end{equation}
where $T_{\mathrm{end}}(q)$ denotes the latency of the plaintext
prefix executed on the end device, $T_{\mathrm{enc}}(q)$ denotes the cost of
encrypting the split activation before offloading,
$T_{\mathrm{comm}}(q,e)$ denotes the aggregate transmission delay across the
active links, 
$T_{\mathrm{edge}}(q,e)$ denotes the protected execution time
incurred at the edge server, 
and $T_{\mathrm{cloud}}(e)$ denotes the additional delay introduced when the cloud completes the remaining suffix, with $T_{\mathrm{cloud}}(M)=0$ in terminate mode. 
Final-logit decryption is omitted from the optimization model
because it is identical for every split candidate; it is included in the
reported absolute end-to-end latency.
In the detailed model below, $T_{\mathrm{comm}}(q,e)$ is
expanded into delays over the active end-to-edge, edge-to-cloud, cloud-to-end, or
edge-to-end links. 

Consequently, the system aims to identify the feasible execution configuration that minimizes end-to-end latency while preserving functional correctness, satisfying the selected split granularity, and maintaining the privacy boundary defined by the system model.
Formally, the optimization objective is:
\begin{equation}
\label{eq:problem-objective}
(q^{*},e^{*}) = \operatorname*{arg\,min}_{(q,e)\in\Omega} T_{\mathrm{tot}}(q,e).
\end{equation}
The feasible set $\Omega$ satisfies four classes of
constraints, including a deployment-specific model-exposure policy.
\begin{itemize}
\item \textbf{Structural constraints.} The partitioned execution order must
remain consistent with the original layer sequence $\mathcal{L}$, ensuring
that the distributed inference path is equivalent to the intact CNN.
\item \textbf{Split-granularity constraints.} Candidate boundaries must
follow the selected search granularity, either allowing convolution-layer
boundaries inside a convolutional block or restricting boundaries to
complete blocks and fully connected layers.
\item \textbf{Privacy constraints.} Raw inputs, plaintext intermediate
representations, and the decryption credential must remain confined to
the trusted end device; only protected intermediate states and protected
outputs may leave the trusted domain. 
\item \textbf{Model-exposure constraints.} The cumulative
fraction of model parameters deployed on the end device must satisfy
\begin{equation}
\label{eq:model-exposure}
W_{\mathrm{end}}(q)\leq W_{\max},
\end{equation}
where $W_{\mathrm{end}}(q)$ is the fraction of total model
parameters from the input through $\ell_q$, and $W_{\max}$ is a policy limit set by
the cloud model owner to reduce model-exposure risk.
\end{itemize}

\section{Latency-Optimal Adaptive Split Inference} \label{sec:split_inference}

\subsection{FHE-Compatible Stage}
As discussed in Section~\ref{sec:prelim_ckks}, CKKS requires
comparison-based nonlinearities to be replaced by arithmetic-compatible
alternatives. We therefore replace each ReLU activation with the
quadratic residual activation:

\begin{equation}
\label{eq:quadratic-residual-activation}
\sigma(z)=(\alpha\cdot z)^{2}+z,
\end{equation}
where $\alpha$ is a learnable scaling coefficient determined during
training. This activation preserves a linear residual pathway alongside
a degree-2 polynomial term, which improves gradient flow compared with
a pure square mapping while maintaining the same CKKS-friendly
multiplicative structure. 

Homomorphic evaluation of this activation requires one
ciphertext multiplication and one addition, keeping multiplicative-depth
growth predictable. The linear operators, including convolution, batch
normalization, and fully connected projection, retain their original
algebraic form. The model is trained and deployed under this
FHE-compatible activation from initialization.

\subsection{Latency-Aware Model Partition Strategy}
Prior to serving any inference request, the system
executes an offline planning phase. 
For latency planning, the ordered stage representation is recalled as
$f(\mathbf{x};\Theta)=f_{M}\circ f_{M-1}\circ\cdots\circ f_{1}(\mathbf{x})$.
Let $h_0=\mathbf{x}$ and $h_m=f_m(h_{m-1})$ denote the output
representation after stage $\ell_m$.
For each stage $\ell_m$, a single forward profiling pass records the
activation cardinality $A_m$ of $h_m$ and the floating-point operation
count $F_m$ required to evaluate $f_m$ for one input sample.
The same checkpoint-specific profiling pass also records the
parameter count of each stage and derives the cumulative end-side model
exposure $W_{\mathrm{end}}(q)$. The profiler output is tied to the exact
checkpoint and input shape used by the encrypted inference pipeline.
These per-stage statistics, together with the end-device compute
rate, benchmark-calibrated remote FHE cost densities, inter-tier
bandwidths, and CKKS configuration, form the planner input.
\runinhead{Cost Model} 
For a candidate split pair $(q,e)\in\Omega$ with $\ell_{q}, \ell_{e}\in \mathcal{L}$ and $e>q$, the latency model is decomposed according to the three execution tiers.

For remote FHE execution, stages with the same encrypted
execution pattern are assigned to a stage-cost group $G$. Let
$F_G=\sum_{m\in G}F_m$ denote the profiled FLOP count of that group for
one input sample, and let $\mathcal{G}_{a:b}$ denote the stage-cost groups
required to execute stages $\ell_a$ through $\ell_b$. For each remote
tier $d\in\{\mathrm{edge},\mathrm{cloud}\}$, an offline benchmark
measures the wall-clock completion time $T^{\mathrm{wall}}_{d,G}$ for
$N^{\mathrm{cal}}_{d,G}$ calibration samples under a fixed cryptographic
and execution configuration. The amortized benchmark latency and the
corresponding empirical FHE cost density are defined as
\begin{equation}
\label{eq:fhe-cost-density}
\widetilde{T}^{\mathrm{bench}}_{d,G}
=\frac{T^{\mathrm{wall}}_{d,G}}{N^{\mathrm{cal}}_{d,G}},
\qquad
\beta_{d,G}
=\frac{\widetilde{T}^{\mathrm{bench}}_{d,G}}{F_G/10^9}.
\end{equation}
Accordingly, the planner estimates the FHE execution cost of
group $G$ on tier $d$ as
\begin{equation}
\label{eq:grouped-he-cost}
T_{d}^{\mathrm{FHE}}(G)
=\beta_{d,G}\frac{F_G}{10^9}.
\end{equation}
The plaintext FLOP count serves only as a stage-size
normalizer. The empirical density $\beta_{d,G}$ absorbs the effects of
ciphertext arithmetic, packing, rotations, kernel implementation,
parallel execution, and hardware. If an auxiliary operation has nonzero
encrypted execution time but no profiled FLOP count, its measured time is
included in the associated stage-cost group. The densities must therefore
be recalibrated whenever the hardware, cryptographic configuration,
packing strategy, kernel implementation, or parallel execution setting
changes. A tier-specific density is calibrated for every supported group,
so all supported stage families remain available to the candidate
search.

\begin{itemize}
    \item \textbf{End device.} The end device executes the plaintext
    prefix from $\ell_1$ to $\ell_q$. Its local computation latency is
    \begin{equation}
    \label{eq:end-latency}
    T_{\mathrm{end}}(q)=\frac{\sum_{m=1}^{q}F_{m}}{R_{\mathrm{end}}\times 10^{9}},
    \end{equation}
    where $R_{\mathrm{end}}$ is the end device's compute rate in GFLOPS.
    
    The split activation at stage $\ell_q$ contains $A_{q}$ scalar values
    per sample. Under CKKS parameterized by ring dimension $N$,
    multiplicative depth $D$, scaling modulus size $b$ bits, and
    sample-interleaved batch factor $B$, each ciphertext occupies
    $C_{\mathrm{ct}}=2N\lceil(D+1)b/64\rceil\times 8$ bytes.
    Since one ciphertext packs $B$ interleaved samples, each
    sample has $\lfloor S/B\rfloor$ usable feature slots. For reference,
    dense packing would require at least
    $\lceil A_m/\lfloor S/B\rfloor\rceil$ ciphertexts per packed batch.
    Channel grouping, padding, and other layout constraints may increase
    the actual count. The layout profiler therefore records
    $n_m^{\mathrm{ct}}$, the actual number of ciphertexts required to
    represent a packed batch at stage $\ell_m$. The corresponding
    amortized ciphertext count and payload per sample are
    \begin{equation}
    \label{eq:profiled-ciphertext-payload}
    K_m=\frac{n_m^{\mathrm{ct}}}{B},
    \qquad
    P_m=\frac{n_m^{\mathrm{ct}}C_{\mathrm{ct}}}{B}.
    \end{equation}
    Accordingly, the encryption latency at split point $q$ is
    \begin{equation}
    \label{eq:encryption-latency}
    T_{\mathrm{enc}}(q)
    =\frac{n_q^{\mathrm{ct}}}{B}\tau_{\mathrm{enc}},
    \end{equation}
    where $\tau_{\mathrm{enc}}$ is an empirically calibrated
    per-ciphertext cost. Transmission of the split activation to the edge
    incurs
    \begin{equation}
    \label{eq:end-edge-comm}
    T_{\mathrm{end}\to\mathrm{edge}}(q)
    =\frac{n_q^{\mathrm{ct}}C_{\mathrm{ct}}}
    {B\,B_{\mathrm{end}\to\mathrm{edge}}},
    \end{equation}
    where $B_{\mathrm{end}\to\mathrm{edge}}$ is the effective end-to-edge uplink bandwidth.
    All payload variables $P$ are measured in amortized bytes
    per sample. Let
    $\overline{B}_{i\to j}$ denote a configured link rate reported in
    Mbps. Before evaluating the communication terms, the planner converts
    it to $B_{i\to j}=10^{6}\overline{B}_{i\to j}/8$ bytes/s. Therefore,
    every communication term of the form $P/B_{i\to j}$ is equivalently
    \begin{equation}
    \label{eq:communication-unit-conversion}
    T_{i\to j}(P)=\frac{P}{B_{i\to j}}
    =\frac{8P}{10^{6}\overline{B}_{i\to j}}.
    \end{equation}
    This model accounts for payload serialization time only;
    fixed propagation delay, queueing delay, and network jitter are not
    included.

    \item \textbf{Edge server.} After receiving the encrypted activation,
    the edge server executes stages $\ell_{q+1}$ through
    $\ell_e$ in the ciphertext domain. Using the empirical
    densities defined above, its FHE execution latency is
    \begin{equation}
    \label{eq:edge-latency}
    T_{\mathrm{edge}}(q,e)=
    \sum_{G\in\mathcal{G}_{q+1:e}}\beta_{\mathrm{edge},G}
    \frac{F_G}{10^9}.
    \end{equation}
    If the edge is not the terminal executor, the ciphertext after layer
    $\ell_e$ is forwarded to the cloud with delay
    \begin{equation}
    \label{eq:edge-cloud-comm}
    T_{\mathrm{edge}\to\mathrm{cloud}}(e)
    =\frac{n_e^{\mathrm{ct}}C_{\mathrm{ct}}}
    {B\,B_{\mathrm{edge}\to\mathrm{cloud}}},
    \end{equation}
    where $B_{\mathrm{edge}\to\mathrm{cloud}}$ is the edge-to-cloud link bandwidth. In the
    \emph{terminate mode} ($e=M$), the edge returns the final encrypted
    logits to the end device with
    \begin{equation}
    \label{eq:edge-end-return}
    T_{\mathrm{edge}\to\mathrm{end}}(M)
    =\frac{n_M^{\mathrm{ct}}C_{\mathrm{ct}}}
    {B\,B_{\mathrm{edge}\to\mathrm{end}}},
    \end{equation}
    where $B_{\mathrm{edge}\to\mathrm{end}}$ is the effective
    bandwidth of the return link from edge to the end device.

    \item \textbf{Cloud server.} The cloud server is used only in the
    relay mode. It completes the residual encrypted suffix from
    $\ell_{e+1}$ to $\ell_M$. Its FHE execution latency is
    \begin{equation}
    \label{eq:cloud-latency}
    T_{\mathrm{cloud}}(e)=
    \sum_{G\in\mathcal{G}_{e+1:M}}\beta_{\mathrm{cloud},G}
    \frac{F_G}{10^9}.
    \end{equation}
    After the cloud obtains the encrypted logits, the return delay to the end
    device is
    \begin{equation}
    \label{eq:cloud-end-return}
    T_{\mathrm{cloud}\to\mathrm{end}}(M)
    =\frac{n_M^{\mathrm{ct}}C_{\mathrm{ct}}}
    {B\,B_{\mathrm{cloud}\to\mathrm{end}}},
    \end{equation}
    where $B_{\mathrm{cloud}\to\mathrm{end}}$ is the effective bandwidth
    of the cloud-to-end return link.
\end{itemize}

\runinhead{Routing Modes and Total Latency} The planner evaluates two
routing modes for every candidate tuple $(q,e)$. In the
\emph{relay mode} ($e<M$), the edge executes stages
$\ell_{q+1}$ through $\ell_{e}$ homomorphically and
forwards the resulting encrypted state to the cloud, which completes the
residual suffix $\ell_{e+1}$ through $\ell_{M}$. The total
latency is

\begin{equation}
\label{eq:relay-latency}
\begin{aligned}
T_{\mathrm{relay}}(q,e)=&
T_{\mathrm{end}}(q)+T_{\mathrm{enc}}(q)+T_{\mathrm{end}\to\mathrm{edge}}(q)
+T_{\mathrm{edge}}(q,e) \\
&+T_{\mathrm{edge}\to\mathrm{cloud}}(e)+T_{\mathrm{cloud}}(e)
+T_{\mathrm{cloud}\to\mathrm{end}}(M).
\end{aligned}
\end{equation}

In the \emph{terminate mode} ($e=M$), the edge completes the full
remaining suffix and returns the encrypted logits directly to the end
device, bypassing the cloud entirely:

\begin{equation}
\label{eq:terminate-latency}
\begin{aligned}
T_{\mathrm{term}}(q,M)=&
T_{\mathrm{end}}(q)+T_{\mathrm{enc}}(q)+T_{\mathrm{end}\to\mathrm{edge}}(q)\\
&+T_{\mathrm{edge}}(q,M)+T_{\mathrm{edge}\to\mathrm{end}}(M).
\end{aligned}
\end{equation}

Accordingly, $T_{\mathrm{tot}}(q,e)$ denotes the candidate end-to-end
latency instantiated by $T_{\mathrm{relay}}(q,e)$ in relay mode and by
$T_{\mathrm{term}}(q,M)$ in terminate mode.

\runinhead{Split Selection} The planner enumerates candidate split pairs
$(q,e)\in\Omega$ with $\ell_q,\ell_e\in\mathcal{L}$ and $e>q$, evaluates
the end-to-end latency of each candidate, and selects the latency-minimizing
deployment plan. Equivalently, the split-selection procedure implements
\[
\pi^{*}=(q^{*},e^{*})=\operatorname*{arg\,min}_{(q,e)\in\Omega}T_{\mathrm{tot}}(q,e).
\]
Two split granularities are considered: convolution-level splitting
allows boundaries inside a convolutional block at convolution-layer granularity,
whereas block-level splitting allows boundaries only between complete blocks or
FC layers and does not split inside a block or an FC layer. 

The complete procedure is given in Algorithm \ref{alg:partition}.

\begin{algorithm}[tb!]
\caption{Offline Latency-Aware Split Selection}
\label{alg:partition}
\begin{algorithmic}[1]
\Require CNN stage sequence $\mathcal{L} = (\ell_1, \ell_2, \ldots, \ell_M)$;
         feasible split-pair set $\Omega$
\Require Profiled stage FLOPs $F_m$ and layout-specific
         ciphertext counts $n_m^{\mathrm{ct}}$
\Require End-device compute rate $R_{\mathrm{end}}$
\Require Configured link rates $\overline{B}_{i\to j}$ in Mbps
\Require CKKS and planner parameters
         $\tau_{\mathrm{enc}},B,N,D,b$; benchmark-calibrated FHE
         densities $\beta_{d,G}$
\Ensure Optimal deployment plan $\pi^* = (q^*, e^*)$

\State Compute per-ciphertext byte size:
       $C_{\mathrm{ct}} \gets 2N \left\lceil (D+1)b / 64 \right\rceil \times 8$
\State Convert each configured link rate:
       $B_{i\to j}\gets 10^6\overline{B}_{i\to j}/8$ bytes/s
\For{$m=1$ to $M$}
    \State $K_m\gets n_m^{\mathrm{ct}}/B$;\quad
           $P_m\gets K_mC_{\mathrm{ct}}$
\EndFor
\State $T_{\mathrm{best}} \gets +\infty$;\quad $\pi^* \gets \emptyset$

\For{each candidate split pair $(q,e)\in\Omega$}
    \State $T_{\mathrm{end}}(q) \gets \sum_{m=1}^{q} F_m \,/\, (R_{\mathrm{end}} \times 10^9)$
    \State $T_{\mathrm{enc}}(q) \gets K_q\tau_{\mathrm{enc}}$;\quad
           $T_{\mathrm{end}\to\mathrm{edge}}(q) \gets
           P_q/B_{\mathrm{end}\to\mathrm{edge}}$
    \State $T_{\mathrm{edge}}(q,e) \gets
           \sum_{G\in\mathcal{G}_{q+1:e}}
           \beta_{\mathrm{edge},G}F_G/10^9$
    \If{$e<M$} \Comment{relay mode: edge $\to$ cloud}
        \State $T_{\mathrm{edge}\to\mathrm{cloud}}(e) \gets
               P_e/B_{\mathrm{edge}\to\mathrm{cloud}}$
        \State $T_{\mathrm{cloud}}(e) \gets
               \sum_{G\in\mathcal{G}_{e+1:M}}
               \beta_{\mathrm{cloud},G}F_G/10^9$
        \State $T_{\mathrm{cloud}\to\mathrm{end}}(M) \gets
               P_M/B_{\mathrm{cloud}\to\mathrm{end}}$
        \State $T_{\mathrm{tot}} \gets T_{\mathrm{end}}(q)+T_{\mathrm{enc}}(q)
               +T_{\mathrm{end}\to\mathrm{edge}}(q)+T_{\mathrm{edge}}(q,e)$
        \Statex \hspace{\algorithmicindent}$\quad
               +T_{\mathrm{edge}\to\mathrm{cloud}}(e)+T_{\mathrm{cloud}}(e)
               +T_{\mathrm{cloud}\to\mathrm{end}}(M)$
    \Else \Comment{terminate mode: edge $\to$ end}
        \State $T_{\mathrm{edge}\to\mathrm{end}}(M) \gets
               P_M/B_{\mathrm{edge}\to\mathrm{end}}$
        \State $T_{\mathrm{tot}} \gets T_{\mathrm{end}}(q)+T_{\mathrm{enc}}(q)
               +T_{\mathrm{end}\to\mathrm{edge}}(q)$
        \Statex \hspace{\algorithmicindent}$\quad
               +T_{\mathrm{edge}}(q,M)+T_{\mathrm{edge}\to\mathrm{end}}(M)$
    \EndIf
    \If{$T_{\mathrm{tot}}<T_{\mathrm{best}}$}
        \State $T_{\mathrm{best}}\gets T_{\mathrm{tot}}$;\quad
               $\pi^*\gets(q,e)$
    \EndIf
\EndFor
\State \Return $\pi^*$

\end{algorithmic}
\end{algorithm}

\subsection{End-Edge-Cloud Collaborative Inference}

Once the optimal partition plan $\pi^{*}$ has been determined offline,
each packed inference batch is processed through an online pipeline that
maintains stage-consistent execution semantics across the three tiers
while enforcing the privacy boundary at the end device.

\runinhead{Local Prefix Inference} The input batch is first processed
through the plaintext prefix of the CNN from the initial layer up to
the selected split point $q^{*}$. This produces the intermediate
activation tensors $h_{q^{*}}$ for the packed samples on the trusted end
device.

\runinhead{Encrypted Feature Packaging} The intermediate activation
tensors $h_{q^{*}}$ are encrypted using a layout-aware packing strategy
that preserves the structural properties of the split-point
representation. When the split occurs within the convolutional backbone,
the activation retains its spatial dimensions and is packed into a
channel-grouped spatial layout in which each ciphertext holds
multi-channel spatial feature elements interleaved across $B$ samples.
When the split occurs after the transition to the classifier head, the
activation is a one-dimensional feature vector and is packed into a
feature-major layout with sample-interleaved slots. In both cases, the
resulting ciphertext sequence $\mathbf{C}_{q^{*}}$ is transmitted
to the edge server over the end-to-edge uplink. The end
device retains the secret key throughout and does not transmit any
plaintext representation of the intermediate features.

\runinhead{Edge-Side Encrypted Continuation} Upon receiving
$\mathbf{C}_{q^{*}}$, the edge server resumes the CNN from layer
$\ell_{q^{*}+1}$ and executes all assigned stages up to the edge
termination point $e^{*}$ entirely within the ciphertext domain.

For convolutional stages, the edge applies BN-fused convolution followed
by quadratic residual activation over the spatial ciphertext layout.
Pooling transitions are realized as compact average-pooling operations,
and the final pooling--flattening transition converts the spatial state
into a feature-major layout. Intermediate fully connected stages evaluate
the encrypted linear projection in the feature-major layout, followed by
BN and the quadratic residual activation. The final classifier stage is
evaluated as a linear projection and produces encrypted logits without
an activation.

Upon completing all assigned stages, the edge either forwards the
resulting encrypted intermediate output to the cloud under the relay
mode or returns the final encrypted logits directly to the end device
under the terminate mode.

\runinhead{Cloud-Side Suffix Completion} In the relay mode, the cloud
receives the encrypted state produced by the edge at layer
$\ell_{e^{*}}$ and completes the residual suffix from
$\ell_{e^{*}+1}$ to $\ell_{M}$ using the same homomorphic computation
pattern. The cloud operates exclusively on ciphertext-domain data and
does not require access to plaintext inputs, intermediate features, or
the secret key.
\runinhead{Local Decryption with Prediction Recovery} The encrypted
logits are returned to the end device, which applies the secret key to
recover the plaintext logit tensor. For sample $r$, the predicted class
is $\widehat{c}_r=\operatorname*{arg\,max}_{j}\widehat{y}_{r,j}$, where
$\widehat{y}_{r,j}$ is its $j$-th decrypted logit.

\begin{figure}[H]
\centering
\includegraphics[width=0.8\textwidth]{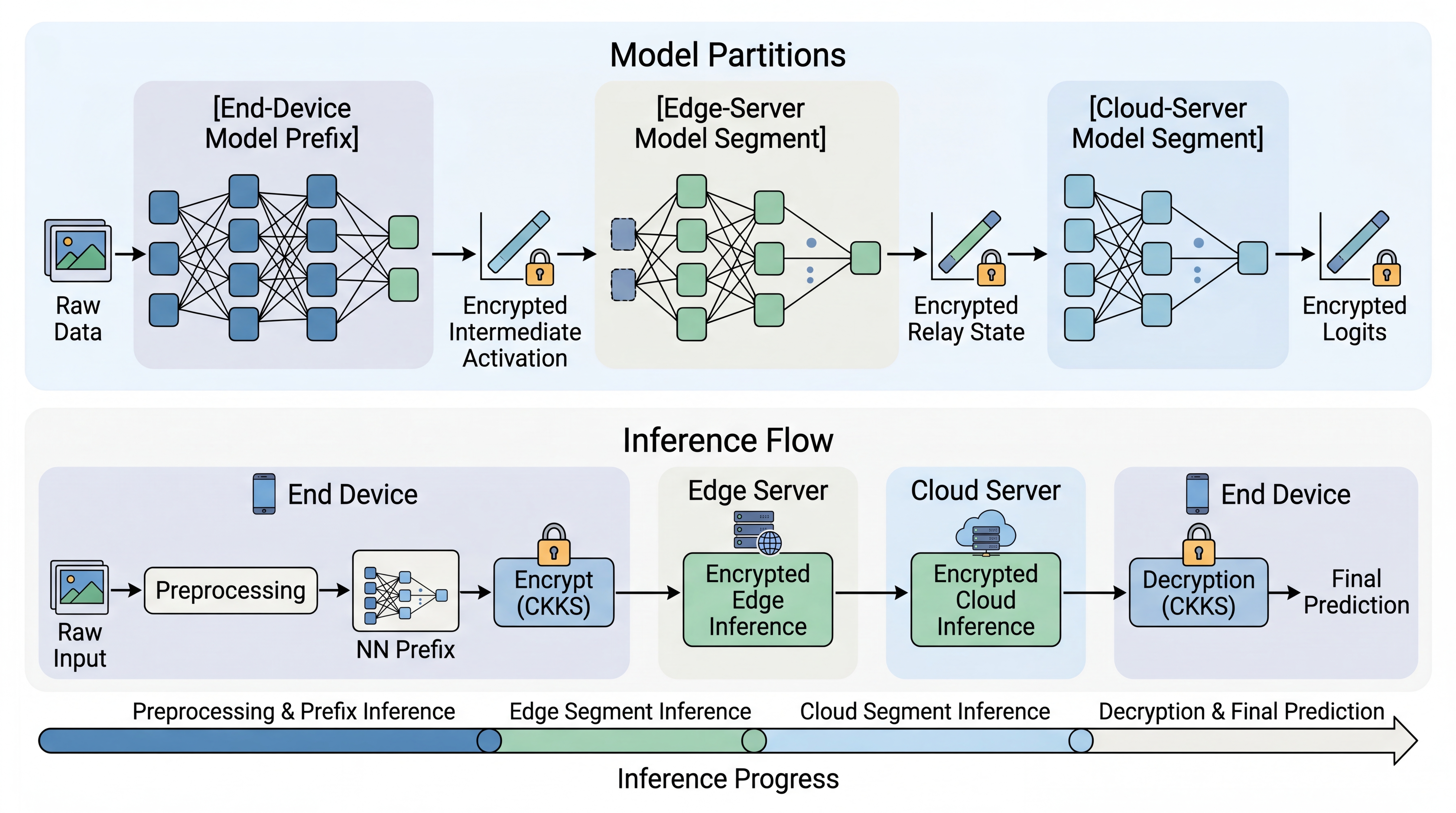}
\caption{Model partitioning and online inference flow of the proposed
privacy-preserving cloud-edge-end split inference architecture.}
\label{fig:ceec_workflow_overview}
\end{figure}

Figure~\ref{fig:ceec_workflow_overview} summarizes the model partitioning
and online inference flow across the end, edge, and cloud tiers.

\section{EXPERIMENTAL PERFORMANCE}

This section reports the experimental protocol, partition
results, and encrypted-inference performance.

\subsection{Experimental Settings}

\runinhead{Hardware Testbed}
The experiments are conducted on a hybrid
physical-and-emulated cloud-edge-end testbed reflecting the architecture
described in Section~\ref{sec:system_model}. The end tier is a separate
physical Raspberry Pi 4B running Ubuntu 22.04 and represents a
resource-constrained trusted endpoint. The edge and cloud tiers are
resource-isolated Docker containers hosted on the same physical server,
which is equipped with a 96-core Intel Xeon Gold 6530 processor at
2.1~GHz. The edge container is allocated 14 CPU cores and 112~GiB of
RAM, whereas
the cloud container is allocated 56 CPU cores and 448~GB of RAM. These
allocations emulate compute-capacity asymmetry between the two remote
tiers; they do not represent two geographically or physically separate
servers. The observed edge-container peak memory usage is 46.6~GiB.
Homomorphic kernel
execution times are measured under the corresponding container resource
limits. The bandwidth values in
Table~\ref{tab:planner_params} are configured effective link rates used
by the planner rather than measurements from wide-area network traces.
Consequently, the testbed does not reproduce geographic propagation
delay, network jitter, cross-region routing, or independently operated
edge and cloud virtualization stacks.

\runinhead{Software Environment}
All experiments are implemented in Python~3.8 with PyTorch~\cite{pytorch}
as the deep learning framework. Homomorphic encryption is realized using the
OpenFHE library~\cite{openfhe} under the CKKS scheme. Parallelism at each tier is
managed via Python multiprocessing.

\runinhead{Model and Dataset}
The underlying classifier follows a VGG16-BN backbone~\cite{vgg} at
$224\times224$ input resolution. Table~\ref{tab:exp_stage_config}
reports the stage-level structural statistics of this backbone and
serves as the profiling reference for subsequent partition planning. The
FHE-compatible SquareVGG16 used in the secure pipeline is derived
from this backbone and is described in detail in the experiment design
subsection below. Two benchmark datasets are used in the experiments:
CIFAR-10~\cite{cifar} (10 classes, 10,000 test images) and PathMNIST~\cite{medmnistv2}
(9 classes, 7,180 test images). All input images are resized to
$224\times224$ pixels and normalized to the respective dataset channel
statistics.

\begin{table}[H]
\caption{Stage-Level Configuration of the Underlying VGG16-BN Backbone.
FLOPs are reported per input sample at $224\times224$ resolution and
follow the convolutional/linear profiling rule used by the partition
planner. Activation size gives the number of scalar elements in the
stage output.
}
\label{tab:exp_stage_config}
\centering
\small
\setlength{\tabcolsep}{3pt}
\resizebox{\textwidth}{!}{%
\begin{tabular}{|l|l|r|r|r|}
\hline
\textbf{Stages} & \textbf{Layer Composition$^{\dagger}$} & \textbf{Output Shape}
  & \textbf{FLOPs (M)} & \textbf{Activation Size}
  \\
\hline
Block I & 2$\times$ Conv 3$\times$3 + BN + Act. + Pool & $64\times112\times112$  & 3872.8  & 802,816  \\
\hline
Block II & 2$\times$ Conv 3$\times$3 + BN + Act. + Pool & $128\times56\times56$  & 5549.1  & 401,408   \\
\hline
Block III & 3$\times$ Conv 3$\times$3 + BN + Act. + Pool & $256\times28\times28$  & 9248.4  & 200,704  \\
\hline
Block IV & 3$\times$ Conv 3$\times$3 + BN + Act. + Pool & $512\times14\times14$  & 9248.4 & 100,352   \\
\hline
Block V & 3$\times$ Conv 3$\times$3 + BN + Act. + Pool & $512\times7\times7$   & 2774.5 &  25,088  \\
\hline
FC1 & FC + BN + Act. & $4096$  & 205.5 & 4,096 \\
\hline
FC2 & FC + BN + Act. & $4096$  & 33.6  & 4,096    \\
\hline
FC3 & Output FC ($C$ classes) & $C$    & 0.08  & $C$     \\
\hline
\multicolumn{5}{l}{\footnotesize $^{\dagger}$Act.: quadratic residual activation; Pool: average pooling; $C=10$ for CIFAR-10 and $C=9$ for PathMNIST.}\\
\end{tabular}
}
\end{table}

\subsection{Experiment Design}

The evaluation proceeds in three phases: FHE-compatible model training,
offline partition planning, and online encrypted inference. This
subsection also defines the baselines used for comparison.

\runinhead{Cryptographic Configuration} All encrypted inference
experiments use the fixed CKKS configuration summarized in
Table~\ref{tab:ckks_params}. The instantiated configuration satisfies
128-bit classical security under the homomorphic encryption security
guidelines~\cite{heguidelines2024}.
The CKKS parameters $N$, $D$, and $b$ are fixed by
multiplicative-depth, security, and numerical-correctness requirements;
they are not latency-fitted cost-model parameters.

\begin{table}[H]
\caption{CKKS Parameter Configuration Used in All Experiments.}
\label{tab:ckks_params}
\centering
\small
\setlength{\tabcolsep}{5pt}
\begin{tabular}{|l|r|l|}
\hline
\textbf{Parameters} & \textbf{Value} & \textbf{Description} \\
\hline
$N$ & 32768 & Ring dimension \\
\hline
$D$ & 7 & Multiplicative depth \\
\hline
$b$ & 50 bits & Scaling modulus size \\
\hline
$S$ & 16384 & SIMD slot count \\
\hline
$B$ & 4 & Sample-interleaved batch factor \\
\hline
$\lfloor S/B \rfloor$ & 4096 & Usable feature slots per sample \\
\hline
\end{tabular}
\end{table}

\runinhead{Phase~I: FHE-Compatible Model Training} All collaborative FHE
experiments use SquareVGG16, the homomorphic variant of the VGG16-BN
backbone described in Section~\ref{sec:split_inference}. The model is trained from scratch under
the FHE-compatible configuration defined in
Section~\ref{sec:split_inference}: the quadratic residual activation in
Eq.~\ref{eq:quadratic-residual-activation}, average pooling, and no
clamping in the encrypted suffix. The scaling coefficient $\alpha$
is fixed at the value recorded in the training checkpoint metadata and
is kept constant throughout inference. Separate checkpoints are trained
for CIFAR-10 and PathMNIST.

\runinhead{Phase~II: Offline Partition Planning}
Before online inference, the planner enumerates the feasible
split pairs and selects $\pi^*$ using Algorithm~\ref{alg:partition}.
Remote FHE cost densities are calibrated under the deployed
cryptographic and execution configuration, and candidate end splits must
satisfy the model owner's $W_{\max}=0.3\%$ exposure policy.

\runinhead{Phase~III: Online Encrypted Inference}
Online measurements follow the planner-selected deployment and
process packed batches of $B=4$ samples. The timing boundary excludes
one-time model loading, CKKS context generation, key generation, and
process startup; it includes the plaintext prefix, encryption, encrypted
stage execution, decryption, and any layout or cache construction on the
online path. Each complete run processes 224 samples in 56 full packed
batches. The edge uses 14 workers over four parallel rounds, whereas the
cloud uses 56 workers in one round. Each FHE worker is restricted to one
OpenMP/BLAS thread, and FHE timing uses critical-path wall-clock
aggregation.

\runinhead{Baseline Configuration} To assess the effectiveness of the
proposed scheme, we compare it against a full-cloud FHE deployment
following the CryptoNet paradigm~\cite{cryptonets}. In this
benchmark-estimated baseline, the end device encrypts 9.25 input
ciphertexts per sample at a cost of
$9.25\tau_{\mathrm{enc}}=5.910$~s/sample. The cloud executes the entire
encrypted inference pipeline, whose latency is assembled from the
benchmark-calibrated stage costs under the same model, CKKS configuration,
and cloud testbed. No partitioning or edge-tier offloading is employed.
A block-level split is also evaluated as a granularity
baseline. It uses the same model, CKKS configuration, and testbed as the
proposed convolution-level plan, but permits split boundaries only
between complete blocks or FC layers.

\subsection{Partition Results}
\label{sec:partition_results}

This subsection reports the planner inputs and selected split
pairs, validates the benchmark-calibrated cost model, and examines the
analytical sensitivity of the selected plan. Table~\ref{tab:planner_params}
lists the cost, policy, layout, and link parameters; the CKKS
configuration appears in Table~\ref{tab:ckks_params}.
The CNN stage sequence operates at individual convolution-layer
granularity, yielding $M=21$ candidate stages that include each
convolutional layer, block-level pooling boundaries, and the three
fully connected layers from FC1 through FC3. This fine-grained
decomposition enables the planner to place split points at
sub-block resolution.

Remote FHE costs use the four measured densities
$\beta_{d,G}$ directly, without nominal edge or cloud GFLOPS rates or
additional multipliers. The configured link rates
$\overline{B}_{i\to j}$ are planner inputs rather than WAN measurements
and are converted to bytes per second according to
Eq.~\ref{eq:communication-unit-conversion}; the resulting communication
terms account for payload serialization only. The parameter $W_{\max}$
is a model-owner policy, and the layout profiler supplies the actual
ciphertext count $n_m^{\mathrm{ct}}$ at every candidate boundary.

\begin{table}[H]
\caption{Planner Input Parameters.}
\label{tab:planner_params}
\centering
\small
\setlength{\tabcolsep}{4pt}
\resizebox{\textwidth}{!}{%
\begin{tabular}{|l|l|r|l|}
\hline
\textbf{Category} & \textbf{Parameters} & \textbf{Value} & \textbf{Description} \\
\hline
Compute & $R_{\mathrm{end}}$ & 1.4 GFLOPS & Configured plaintext-rate normalizer \\
\hline
\multirow{4}{*}{FHE Cost Density}
 & $\beta_{\mathrm{edge},\mathrm{conv+pool}}$
 & 115.982 s/GFLOP
 & Measured edge convolution-plus-pooling density \\
\cline{2-4}
 & $\beta_{\mathrm{edge},\mathrm{FC}}$
 & 256.015 s/GFLOP
 & Measured 14-worker edge FC density \\
\cline{2-4}
 & $\beta_{\mathrm{cloud},\mathrm{conv+pool}}$
 & 40.451 s/GFLOP
 & Measured cloud convolution-plus-pooling density \\
\cline{2-4}
 & $\beta_{\mathrm{cloud},\mathrm{FC}}$
 & 35.684 s/GFLOP
 & Measured cloud FC density \\
\hline
Policy & $W_{\max}$
 & 0.3\%
 & Maximum end-side model exposure \\
\hline
\multirow{4}{*}{Bandwidth} & $\overline{B}_{\mathrm{end}\to\mathrm{edge}}$ & 1000 Mbps & End-to-edge link \\
\cline{2-4}
 & $\overline{B}_{\mathrm{edge}\to\mathrm{cloud}}$ & 10000 Mbps & Edge-to-cloud link \\
\cline{2-4}
 & $\overline{B}_{\mathrm{edge}\to\mathrm{end}}$ & 1000 Mbps & Edge-to-end return link \\
\cline{2-4}
 & $\overline{B}_{\mathrm{cloud}\to\mathrm{end}}$ & 1000 Mbps & Cloud-to-end return link \\
\hline
Planner & $\tau_{\mathrm{enc}}$
 & 0.639 s/ciphertext
 & Measured amortized encryption cost \\
\hline
Layout Profile & $n_m^{\mathrm{ct}}$
 & Stage dependent
 & Actual ciphertext count at each candidate boundary \\
\hline
\end{tabular}
}
\end{table}

\runinhead{Calibration}
The edge benchmark covers the convolution-plus-pooling family.
It measures 214.530~s/sample for a group containing 1.849688064 profiled
GFLOPs, giving the empirical density
\begin{equation}
\label{eq:edge-factor-instance}
\beta_{\mathrm{edge},\mathrm{conv+pool}}
=\frac{214.530}{1.849688064}
\approx115.982\ \mathrm{s/GFLOP}.
\end{equation}
Under the fixed edge configuration described in the
Experimental Settings, the measured FC1 critical-path time
is 2946.516~s for one parallel round of 56 samples, corresponding to an
amortized latency of 52.616~s/sample (CV 0.88\%). The complete FC1--FC3
benchmark measures 61.228~s/sample for 239,157,248 profiled FLOPs,
yielding
\begin{equation}
\label{eq:edge-fc-density}
\beta_{\mathrm{edge},\mathrm{FC}}
=\frac{61.228}{239157248/10^9}
=256.015\ \mathrm{s/GFLOP}.
\end{equation}
The same aggregate FC density is applied to individual FC
stages in proportion to their profiled FLOPs when the planner evaluates
candidate boundaries between FC layers.

\runinhead{Selected Split}
The planner evaluates every feasible pair $(q,e)$ under each
split granularity. Table~\ref{tab:optimal_split} reports the selected
convolution-level plan and the block-level reference plan used in the
ablation.

\begin{table}[H]
\caption{Selected Split Pairs Under Different Split Granularities.}
\label{tab:optimal_split}
\centering
\small
\begin{tabular}{|l|l|l|}
\hline
\textbf{Split Granularity} & \textbf{End-side Split Point $q^*$} & \textbf{Edge-side Split Point $e^*$}  \\
\hline
Convolution-level & Block~II & Conv.~III-1 \\
\hline
Block-level & Block~II & Block~III \\
\hline
\end{tabular}
\end{table}

The exposure policy determines the deepest feasible end boundary.
Block~II requires 260,928 cumulative parameters,
corresponding to 0.194\% of the complete model and 0.995~MiB in FP32,
and therefore satisfies $W_{\max}=0.3\%$. Moving the end split to
Conv.~III-1 would expose 556,608 parameters, or 0.414\% and 2.123~MiB,
and is infeasible under this policy. Relative to Block~I, Block~II reduces the split
activation cardinality from 802,816 to 401,408 elements, a 50\%
reduction. Its deployed layout contains 104 ciphertexts per packed batch,
or 26 ciphertexts per sample at $B=4$; the measured encryption latency
of 16.611~s/sample gives
$\tau_{\mathrm{enc}}=16.611/26=0.639$~s/ciphertext. The two search
granularities then produce the assignments in
Table~\ref{tab:optimal_split}.

\runinhead{Validation}
We validate the cloud cost model on CIFAR-10. The measured
suffix latency is 800.320~s/sample, comprising
519.569~s/sample of convolution (64.92\%), 272.217~s/sample of pooling
(34.01\%), and 8.534~s/sample of FC1--FC3 computation (1.07\%). Because
pooling accounts for 34.01\% of the suffix latency, it is folded into the
preceding convolutional calibration group.
Table~\ref{tab:cost_model_validation} distinguishes the calibration fits
from the held-out validation result.

\begin{table}[H]
\caption{Validation of the Benchmark-Calibrated Cloud Cost Model on CIFAR-10.}
\label{tab:cost_model_validation}
\centering
\small
{
\resizebox{\textwidth}{!}{%
\begin{tabular}{|l|r|r|r|}
\hline
\textbf{Validation Scope} & \textbf{Measured (s/sample)}
 & \textbf{Predicted (s/sample)} & \textbf{Relative Error} \\
\hline
Cloud Block~V (held out) & 118.390 & 112.233 & $-5.20\%$ \\
\hline
Cloud FC1--FC3 (calibration) & 8.534 & 8.534 & $0.00\%$ \\
\hline
Complete cloud suffix & 800.320 & 794.162 & $-0.77\%$ \\
\hline
\end{tabular}
}
}
\end{table}

The cloud convolution-plus-pooling density is calibrated using
Blocks~III--IV, leaving Block~V as an independent held-out group with a
5.20\% prediction error. The cloud FC1--FC3 row is a calibration fit,
and the complete suffix is an aggregate reconstruction check. Likewise,
the measured edge latency of 214.530~s/sample is a calibration input, not
a held-out validation result.

\runinhead{Analytical Deployment Sensitivity}
We assess deployment sensitivity by rerunning the calibrated
planner without repeating FHE execution. For the bandwidth sweep, the
current convolution-family rate ratio is fixed at 0.35 while the
end-to-edge and edge-to-cloud rates vary. For the compute-rate sweep, the
links are fixed at 1000 and 10000~Mbps, respectively; the cloud density
is held fixed and the edge convolution-plus-pooling density is set by the
effective edge/cloud FHE-rate ratio
\[
\rho_{e/c}
=\frac{1/\beta_{\mathrm{edge},\mathrm{conv+pool}}}
{1/\beta_{\mathrm{cloud},\mathrm{conv+pool}}}
=\frac{\beta_{\mathrm{cloud},\mathrm{conv+pool}}}
{\beta_{\mathrm{edge},\mathrm{conv+pool}}}.
\]
All remaining densities, CKKS parameters, and the
$W_{\max}=0.3\%$ policy are held fixed in both sweeps. The bandwidth
panel of Table~\ref{tab:planner_sensitivity} indexes end-to-edge rates
by row and edge-to-cloud rates by column, both in Mbps. All nine
bandwidth combinations select Block~II / Conv.~III-1, while the
analytical objective ranges from 1032.947 to 1122.763~s/sample. The same
split is selected at $\rho_{e/c}=0.10$, 0.35, and 0.70; when the edge
rate reaches the cloud rate, the edge termination moves to Block~V.
These are planner outputs under configured bandwidths, not additional
FHE executions or geographically distributed WAN measurements;
propagation delay and network jitter are not modeled.

At the current configuration, the configured
$R_{\mathrm{end}}$ predicts 6.730~s/sample for the plaintext prefix,
whereas the measured value is 0.657~s/sample; the cloud model predicts
794.162~s/sample versus the measured 800.320~s/sample. The aggregate
analytical objective is 1032.947~s/sample and omits the split-invariant
decryption cost. Because its component errors offset, its proximity to
the corresponding 1033.033~s/sample sum of measured computation and
modeled communication is not treated as independent validation; the
held-out Block~V error is the relevant result.

\begin{table}[tbp]
\caption{Analytical Planner Sensitivity (Objectives in Seconds per Sample).}
\label{tab:planner_sensitivity}
\centering
\small
\begin{minipage}[t]{0.46\textwidth}
\centering
\textbf{(a) Bandwidth Grid}\par\vspace{2pt}
\resizebox{\linewidth}{!}{%
\begin{tabular}{|c|r|r|r|}
\hline
\textbf{End-to-Edge} & \multicolumn{3}{c|}{\textbf{Edge-to-Cloud (Mbps)}} \\
\cline{2-4}
\textbf{(Mbps)} & \textbf{100} & \textbf{1000} & \textbf{10000} \\
\hline
10   & 1122.763 & 1109.815 & 1108.520 \\
100  & 1054.060 & 1041.113 & 1039.818 \\
1000 & 1047.190 & 1034.242 & 1032.947 \\
\hline
\multicolumn{4}{|c|}{All settings: Block~II / Conv.~III-1} \\
\hline
\end{tabular}
}
\end{minipage}\hfill
\begin{minipage}[t]{0.52\textwidth}
\centering
\textbf{(b) Effective FHE-Rate Sweep}\par\vspace{2pt}
\resizebox{\linewidth}{!}{%
\begin{tabular}{|c|l|r|}
\hline
\textbf{$\rho_{e/c}$} & \textbf{Selected $(q^*,e^*)$}
 & \textbf{Objective} \\
\hline
0.10 & \multirow{3}{*}{Block~II / Conv.~III-1} & 1569.272 \\
\cline{1-1}\cline{3-3}
0.35 (current) & & 1032.947 \\
\cline{1-1}\cline{3-3}
0.70 & & 925.682 \\
\hline
1.00 & Block~II / Block~V & 893.101 \\
\hline
\end{tabular}
}
\end{minipage}
\end{table}

\subsection{Inference Results}

Measured latency values are amortized over the 224 samples in
each complete run, whereas classification accuracy is evaluated on the
1000-sample subsets specified below. 
The deployment-sensitivity results
are analytical planner outputs rather than additional measurements.

\runinhead{End-to-End Latency}
Table~\ref{tab:latency_results} combines measured split-plan components
and benchmark-estimated full-cloud execution with communication modeled
from profiled payloads and configured link rates. The last column is
3600 divided by the total amortized latency.

The convolution-level plan requires 1033.279~s/sample on
CIFAR-10 and 1023.429~s/sample on
PathMNIST. These results correspond to speedups of $12.927{\times}$ and
$12.775{\times}$ relative to full-cloud FHE and $3.889{\times}$ and
$3.925{\times}$ relative to the block-level alternative, respectively.
Remote FHE execution remains the dominant cost.

\begin{table}[t!]
\caption{Amortized Latency on CIFAR-10 and PathMNIST. Latencies are in
seconds per sample, throughput is in samples per hour, and communication
is modeled under the configured link rates.}
\label{tab:latency_results}
\centering
\small
\setlength{\tabcolsep}{3pt}
\resizebox{\textwidth}{!}{%
\begin{tabular}{|l|l|r|r|r|r|r|r|r|r|}
\hline
\textbf{Dataset} & \textbf{Configuration} & \textbf{End}
  & \textbf{Encrypt} & \textbf{Edge} & \textbf{Cloud}
  & \textbf{Comm.} & \textbf{Decrypt} & \textbf{Total}
  & \textbf{Throughput} \\
\hline
\multirow{3}{*}{CIFAR-10}
 & Full-cloud FHE & -- & 5.910 & -- & 13350.875 & 0.279 & 0.237 & 13357.301 & 0.270 \\
\cline{2-10}
 & Block-level split & 0.624 & 16.646 & 2753.542 & 1246.901 & 0.807 & 0.247 & 4018.771 & 0.896 \\
\cline{2-10}
 & Convolution-level split & 0.657 & 16.611
 & 214.530 & 800.320 & 0.915 & 0.246
 & 1033.279 & 3.484 \\
\hline
\multirow{3}{*}{PathMNIST}
 & Full-cloud FHE & -- & 5.910 & -- & 13068.130 & 0.279 & 0.231 & 13074.550 & 0.275 \\
\cline{2-10}
 & Block-level split & 0.648 & 16.592 & 2762.980 & 1235.234 & 0.807 & 0.224 & 4016.485 & 0.896 \\
\cline{2-10}
 & Convolution-level split & 0.635 & 16.539
 & 211.490 & 793.621 & 0.915 & 0.229
 & 1023.429 & 3.518 \\
\hline
\end{tabular}
}
\end{table}

Each run contains 224 samples in 56 packed batches; multiplying the
amortized totals by 224 gives 231454.399~s and 229247.999~s, respectively.

\begin{samepage}
\runinhead{Accuracy Under FHE Computation}
Table~\ref{tab:accuracy_results} reports
top-1 accuracy for plaintext standard VGG16 (reference), plaintext
SquareVGG16, and the proposed scheme on both datasets. The accuracy gap
is reported in percentage points (pp)
relative to plaintext standard VGG16. Each reported accuracy is the mean
over ten runs, each evaluating a 1000-sample test subset; within each
run, the compared schemes use the same sampled subset.

The proposed scheme achieves 88.53\% on CIFAR-10, matching the plaintext
SquareVGG16 accuracy. On PathMNIST, the ciphertext-domain accuracy is
91.43\%, differing from the plaintext model by only 0.01~pp, consistent
with CKKS floating-point precision effects. Thus, encrypted execution
introduces no meaningful accuracy degradation on either benchmark.
\end{samepage}

\begin{table}[H]
\caption{Classification Accuracy on CIFAR-10 and PathMNIST.}
\label{tab:accuracy_results}
\centering
\small
\setlength{\tabcolsep}{4pt}
\resizebox{\textwidth}{!}{%
\begin{tabular}{|l|l|l|r|r|}
\hline
\textbf{Dataset} & \textbf{Scheme} & \textbf{Execution Mode}
  & \textbf{Top-1 Acc. (\%)} & \textbf{Gap (pp)} \\
\hline
\multirow{3}{*}{CIFAR-10}
 & Standard VGG16 & Plaintext & 93.40 & -- \\
\cline{2-5}
 & SquareVGG16 & Plaintext & 88.53 & 4.87 \\
\cline{2-5}
 & Proposed collaborative inference
   & Hybrid (end plaintext + remote FHE) & 88.53 & 4.87 \\
\hline
\multirow{3}{*}{PathMNIST}
 & Standard VGG16 & Plaintext & 94.62 & -- \\
\cline{2-5}
 & SquareVGG16 & Plaintext & 91.44 & 3.18 \\
\cline{2-5}
 & Proposed collaborative inference
   & Hybrid (end plaintext + remote FHE) & 91.43 & 3.19 \\
\hline
\end{tabular}
}
\end{table}

\begin{figure}[H]
\centering
\includegraphics[width=\textwidth,height=0.42\textheight,keepaspectratio]{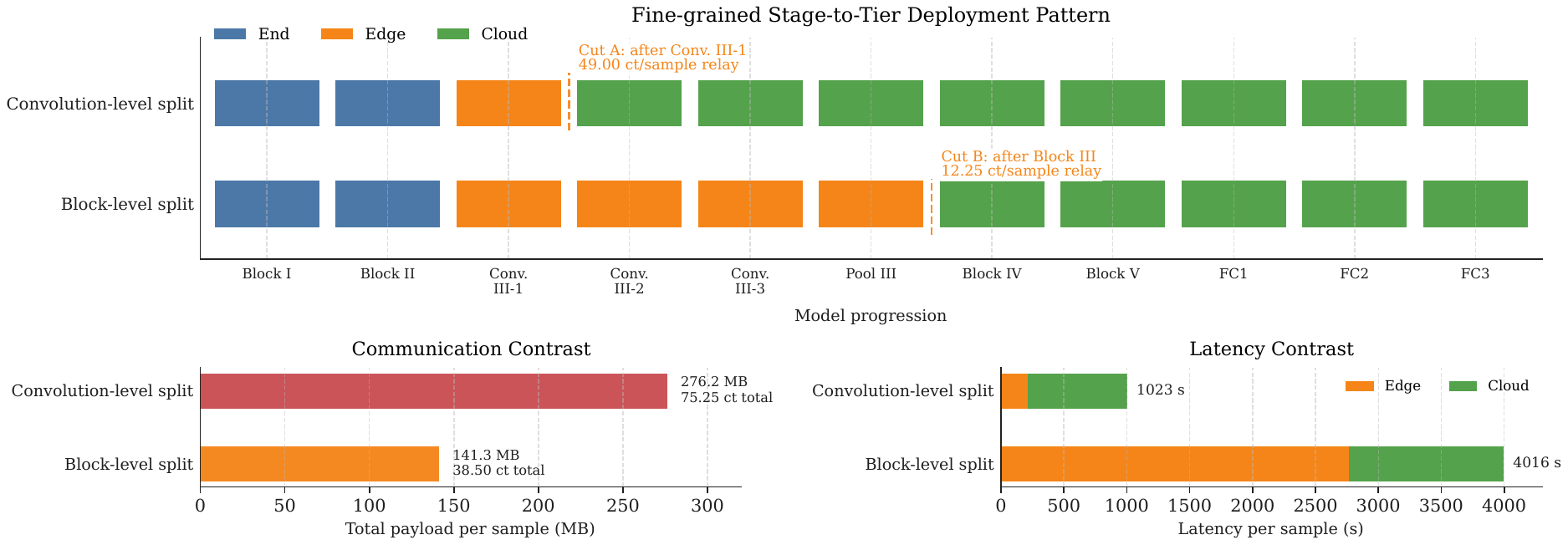}
\caption{Split-plan communication and latency trade-off on
PathMNIST.}
\label{fig:split_visualization_calibrated}
\end{figure}

\runinhead{Communication Overhead}
Table~\ref{tab:communication} reports the total amortized per-sample
ciphertext payload summed over all active inter-tier links. The convolution-level split incurs
approximately $2{\times}$ the communication volume of the block-level
alternative (276.2~MB versus 141.3~MB) because the edge-to-cloud
handoff occurs after Conv.~III-1 ($256{\times}56{\times}56$ elements)
rather than after the pooled Block~III output
($256{\times}28{\times}28$ elements).
Despite this additional ciphertext traffic, the lower
edge-side FHE cost makes the convolution-level plan faster overall. The
full-cloud total comprises 9.25 input and 0.25 output ciphertexts per
sample; only the input ciphertexts contribute to encryption cost.
The
communication-computation trade-off is summarized in Figure~\ref{fig:split_visualization_calibrated} .

\begin{table}[H]
\caption{Amortized Communication Overhead on PathMNIST
($B=4$).}
\label{tab:communication}
\centering
\small
\setlength{\tabcolsep}{4pt}
\resizebox{\textwidth}{!}{%
\begin{tabular}{|l|r|r|}
\hline
\textbf{Configuration} & \textbf{Amortized Ciphertext Blocks / Sample} & \textbf{Payload (MB)} \\
\hline
Full-cloud FHE & 9.50 & 34.865 \\
\hline
Block-level split & 38.50 & 141.296 \\
\hline
	Convolution-level split & 75.25 & 276.169 \\
\hline
\end{tabular}
}
\end{table}

\runinhead{Privacy Protection}
All evaluated configurations protect transmitted states and
remote computation with CKKS. Under the CKKS security assumptions and
the threat model in Section~\ref{sec:system_model}, the external tiers
receive ciphertexts but not the secret key, even if their observations
are combined. The split plans therefore preserve the same cryptographic
privacy boundary as full-cloud FHE; they change only where the encrypted
operators execute. The cryptographic configuration is reported in
Table~\ref{tab:ckks_params}.

\section{CONCLUSION}
This paper presented a systems-level framework for
privacy-preserving split inference in cloud-edge-end environments. The
framework uses established CKKS-based FHE to keep raw inputs, plaintext
intermediate states, and the secret key on the end device, while the edge
and cloud process assigned CNN segments only in the ciphertext domain.
We formulated deployment as a split-pair selection problem and developed
a benchmark-calibrated latency planner that supports convolution-level
and block-level partitioning.

On CIFAR-10 and PathMNIST, the selected convolution-level plan
achieves approximately $12.9{\times}$ speedup over full-cloud FHE and
$3.9{\times}$ over block-level splitting. Its amortized end-to-end
latencies are 1033.279 and 1023.429~s/sample, respectively, including
modeled communication under the configured link rates. The
analytical sensitivity study further shows that the selected split is
stable across the configured bandwidth grid and moves deeper into the
edge only when the effective edge FHE rate approaches the cloud rate.

The current OpenFHE implementation remains a CPU-based research
prototype for non-interactive batch inference rather than real-time
decision making. Evaluation is limited to SquareVGG16 on CIFAR-10 and
PathMNIST and to a hybrid testbed that emulates edge-cloud resource
asymmetry without reproducing a geographically distributed network.
Future work will investigate hardware acceleration, more compact
FHE-compatible architectures, improved ciphertext packing and rotation
schedules, pipeline parallelism, and validation across broader model
families and physically distributed deployments.

\bibliographystyle{splncs04}
\bibliography{references}
\end{document}